\documentclass[aps,prl,twocolumn,showpacs]{revtex4}
\usepackage{graphicx}
\usepackage{color}

\begin{document}

\title{Mechanism of formation of half-doped stripes in underdoped cuprates}

\author{Chung-Pin Chou}
\affiliation{Institute of Physics, Academia Sinica, Nankang, Taiwan}
\author{Ting-Kuo Lee}
\affiliation{Institute of Physics, Academia Sinica, Nankang, Taiwan}

\begin{abstract}
Using a variational Monte-Carlo approach with a recently proposed
stripe wave function, we showed that the strong correlation included
in a $t-J-$type model has essentially all the necessary ingredients
to form these stripes with modulations of charge density, spin
magnetization, and pair field. If a perturbative effect of
electron-phonon coupling to renormalize the effective mass or the
hopping rate of holes is considered with the model, we find the
half-doped stripes, which has on the average one half of a hole in
one period of charge modulation, to be most stable, energetic wise
in the underdoped region, $1/12\leq\delta\leq1/8$. This is in good
agreement with the observation in the neutron scattering
experiments. We also find long range Coulomb interaction to be less
effective in the formation of half-doped stripes.
\end{abstract}

\maketitle

The mechanism for the formation of the stripe states has been one of
the most important issues in understanding high-temperature
superconducting cuprates. Although many theoretical works based on
the electronic model alone have succeeded in comprehending
experimental results qualitatively \cite{AndersonJPC04,PALeeRMP06},
there are many puzzles on the stripe formation in the underdoped
cuprates $La_{2-x}Sr_{x}CuO_{4}$ (LSCO) and $La_{2-x}Ba_{x}CuO_{4}$
(LBCO) with $1/8$ doping most stable
\cite{TranquadaNat95,KYamadaPRB98,MFujitaPRB02,AbbamonteNatPhys05,MitrovicPRB08}.
For instance, the doping dependence of the incommensurate magnetic
peaks associated with the stripe determined by neutron scattering
experiment on LSCO obeys the so-called "Yamada plot"
\cite{KYamadaPRB98} where it points to the existence of the
half-doped stripe with average of $1/2$ hole in one charge
modulation period below $1/8$ hole density. Furthermore, at $1/8$
doping of LBCO, the local density of states still has a V-shape and
a node at low energy \cite{VallaSci06}, even though the
superconductivity is almost completely suppressed by the static
stripe. Surprisingly, a recent scanning tunneling spectroscopy has
observed cluster glasses with randomly oriented domains of
stripe-like patterns in two other families of
$Ca_{2-x}Na_{x}CuO_{2}Cl_{2}$ and
$Bi_{2}Sr_{2}Dy_{0.2}Ca_{0.8}Cu_{2}O_{8+\delta}$
\cite{YKohsakaSci07}. So far all these results are yet to be
explained theoretically in a strongly-correlated framework.

Many early theoretical works have found that stripe-ordered
formation may be present in the extended $t-J$ Hamiltonian
\cite{ZaanenJPCS98,ArrigoniPRB02,HimedaPRL02,WhiteScalapinoPRL}. In
particular, using the density matrix renormalization group (DMRG) in
the standard $t-J$ models, the mechanism of the half-doped stripe
appears to have been understood by antiferromagnetic correlation
across the hole \cite{MartinsPRL00} or optimum energy per hole of a
domain wall \cite{WhitePRL98}. However due to small lattice size,
weak coupling treatment, and special boundary conditions used in
these works, this issue is still being debated \cite{HellbergPRL99}.
The result of DMRG indicated the stripe to be most stable when the
second nearest-neighbor hopping, t', is very small. This result is
exactly opposite to the previous variational Monte Carlo (VMC)
calculation \cite{HimedaPRL02}. Based upon the idea of $d$-wave
resonating-valence-bond ($d$-RVB) state, recently we have proposed a
new stripe-ordered wave function which involves modulations of
charge, spin magnetization, and pair field. These stripe states
either in a periodic pattern or a randomly oriented domain have
almost the same energy as the uniform $d$-RVB state for almost all
values of $t'/t$ in the extended $t-J$ model \cite{CPChouPRB08}.
This unexpected energy degeneracy of stripe-ordered states seems to
be quite compatible with the proposal of fluctuating stripes
\cite{KivelsonRMP03}. However, this degeneracy between different
stripe orders and a uniform state in the $t-J-$ type model fails to
explain the prominence of the half-doped stripes.

There are several possibilities for the formation of stripes.
Long-range Coulomb interaction between holes has been proposed
sometime ago to transform the frustrated phase separation into a
stripe state \cite{EmeryPhysicaC93,JHanJJMPB01}. On the other hand,
a number of experiments have presented evidences that phonon or
lattice effects are apparently present in these materials
\cite{BarYamBook92,FukudaPRB05,ReznikNat06,XJZhouBook07}.
Particularly, due to the structural instability, the isotope effect
in LSCO and LBCO is also peculiarly strong near $1/8$ doping
\cite{CrawfordSci90}. Moreover, to understand a kink structure in
angle-resolved photoemission spectrum \cite{LanzaraNat01}, many
theoretical works have accepted that the electron-phonon couplings
should be considered in the $t-J-$type models
\cite{SzczepanskiZPB92,JLorenzanaPRL95,RoschPRL04,SIshiharaPRB04,ASMishchenkoPRL08}.

In this Letter, we shall consider both long-range Coulomb
interaction and electron-phonon interaction on the formation of
stripes, in particular half-doped stripes, within the context of
strong correlation as in the extended $t-J$ model. However
electron-phonon interaction has many effects. We will only examine
the simplest effect of mass renormalization of charges due to phonon
coupling when the coupling strength is rather weak. The
renormalization effect depending on the local charge density is
treated self-consistently in a VMC method which takes into account
the strong correlation accurately. In the latter half of the paper,
we present results of including long-ranged repulsive interaction
between holes in the $t-J$ Hamiltonian without electron-phonon
interaction.

Let us begin with an extended $t-J$ Hamiltonian in a two-dimensional
square lattice which is defined as,
\begin{eqnarray}
H=-\sum_{i,j,\sigma}t_{ij}\left(\tilde{c}_{i\sigma}^{\dag}\tilde{c}_{j\sigma}+h.c.\right)+J\sum_{<i,j>}\mathbf{S}_{i}\cdot\mathbf{S}_{j}.
\label{e:Equ1}
\end{eqnarray}
The hopping amplitude $t_{ij}=t$, $t'$, and $t''$ for sites $i$ and
$j$ being the nearest-, the second-nearest, and the
third-nearest-neighbors, respectively. Other notations are standard.
In the following, we focus on the case $(t',t'',J)/t=(-0.2,0.1,0.3)$
as we have shown \cite{CPChouPRB08} that comparing with the uniform
state, energies of the stripe state are not very sensitive to the
values of $t'/J$ and $t"/J$. Following our treatment in
Ref.\cite{CPChouPRB08}, the stripe wave function is constructed by
the mean-field Bogoliubov de Gennes method. We assume that the
spatial dependence of the local charge density $\rho_{i}$, staggered
magnetization $m_{i}$, and nearest-neighbor pair field $\Delta_{ij}$
has the simple forms:
\begin{equation}
\rho_{i}=\rho_{v}\cos[2q\cdot(y_{i}-y_{0})], \label{e:Equ2}
\end{equation}
\begin{equation}
m_{i}=m_{v}\sin[q\cdot(y_{i}-y_{0})], \label{e:Equ3}
\end{equation}
\begin{eqnarray}
\Delta_{i,i+\hat{x}}&=&\Delta^{M}_{v}\cos[2q\cdot(y_{i}-y_{0})]-\Delta^{C}_{v},\nonumber\\
\Delta_{i,i+\hat{y}}&=&-\Delta^{M}_{v}\cos[2q\cdot(y_{i}-y_{0})+q]+\Delta^{C}_{v},
\label{e:Equ4}
\end{eqnarray}
where $q=\pi/a_{c}$ and $a_{c}$ is the period of charge modulation.
Here the stripe extends uniformly along $\hat{x}$ direction, and
$y_{0}$ is $1/2$ $(0)$ for the bond-centered (site-centered) stripe.
Sited-centered stripe will not be considered
 throughout the paper unless specifically mentioned. This
state called AF-RVB stripe state \cite{CPChouPRB08} is quite
different from the antiphase stripe studied by others
\cite{HimedaPRL02}. The magnitude of pair field and the
magnetization are larger at sites with smaller hole density. Once
the variational parameters are given, we diagonalize the mean-field
Hamiltonian. Then, the trial wave function with a Gutzwiller
projector $P$ can be constructed by creating all negative energy
states $\bar{\gamma}_{n}^{\dag}$ and annihilating all positive
energy states $\gamma_{n}$ on a vacuum of electrons $|0\rangle$,
\begin{eqnarray}
PP_{J}P_{N_{e}}\prod_{n}\gamma_{n}\bar{\gamma}^{\dag}_{n}|0\rangle,
\label{e:Equ5}
\end{eqnarray}
with the fixed number of electrons $N_{e}$. Here we introduce a
particle-hole transformation \cite{HYokoyamaJPSJ88} in
Eq.(\ref{e:Equ5}), $c_{i\uparrow}^{\dag}\rightarrow f_{i}$ and
$c_{i\downarrow}^{\dag}\rightarrow d_{i}^{\dag}$. Using this
transformation, we avoid the possibility of having a divergent
determinant because of the presence of nodes in RVB-type wave
function with the fixed number of particles using periodic boundary
condition. We also introduce a hole-hole repulsive Jastrow factors
$P_{J}$ in order to greatly improve the variational energies (See
the details in Ref.\cite{CPChouPRB08}).

\begin{figure}[top]
\rotatebox{0}{\includegraphics[height=1.75in,width=2.5in]{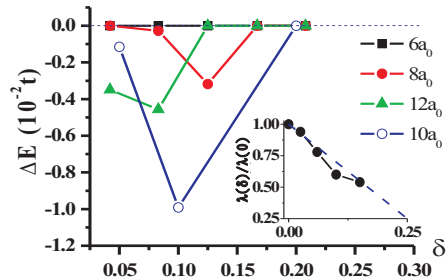}}
\caption{(Color online) The optimized energy difference between the
AF-RVB stripe and uniform $d$-RVB states vs the doping density
$\delta$ for different periods of stripe. $10a_{0}$ is calculated in
$20\times20$ lattice and others in $24\times24$ one. Here $\Lambda$
is  $0.25$. \textbf{Inset:} the electron-phonon coupling constant vs
hole density. Filled circles show the data points from
Ref.\cite{ASMishchenkoPRL08}. The dashed line is $1-3\delta$.}
\label{f:Fig1}
\end{figure}

The simplest effect of an electron-phonon interaction is to
renormalize the mass of charge carriers. To take into account of it
within the extended $t-J$ model, we assume the hopping terms
$t_{ij}$ in Eq.(\ref{e:Equ1}), which are inversely proportional to
the effective mass, are modified due to the spatial variation of
hole density. We neglect the renormalization of the exchange
interaction $J$ since it is only dominant in the very underdoped
region with a very small effect \cite{JLorenzanaPRL95}. We also do
not consider the on-site energy that may arise from the
electron-phonon coupling \cite{KaneshitaPRB07}. If the on-site
energy varies linearly with the local hole density, it will have the
same energy irrespective of the period of periodic stripes.
According to the experimental results that the electron-phonon
coupling strength $\lambda$ is reduced by the hole density $\delta$
\cite{ASMishchenkoPRL08}, we assume a linear relation between
$\lambda$ and $\delta$, $\lambda(\delta)/\lambda(0)\equiv
f(\delta)=1-3\delta$ (See the inset in Fig.\ref{f:Fig1}).
Accordingly, sites with a larger hole density in a modulated charge
pattern have larger $t_{ij}$ \cite{note}. Thus $t_{ij}$ in
Eq.(\ref{e:Equ1}) is renormalized to $t_{ij}^{\ast}$, given by
\begin{equation}
t_{ij}^{\ast}=t_{ij}\left(1-\Lambda\left(\frac{f(n_{h}^{i})+f(n_{h}^{j})}{2}\right)\right),
\label{e:Equ6}
\end{equation}
where $n_{h}^{i}$ is the hole density at site $i$ and
$\Lambda=\lambda(0)/\pi$. We estimate a reasonable range of
$\Lambda$ should be from $0$ to $0.25$ \cite{note}. Since $\lambda$
depends on the hole density $n_{h}^{i}$, we have to do a
self-consistent calculation to find the minimal variational energy.
Here we use the iterative method to achieve the convergence. First
we assume an initial set of values for the hole density at different
sites such that $t_{ij}^{\ast}$ are given, then we can proceed to
optimize the trial wave functions. If the hole densities calculated
by this optimized wave function do not agree with the initial input
values, we will then use the calculated density, hence its
$t_{ij}^{\ast}$, as input value for the next round of optimization.
This process is repeated until we have a converged result. Sometimes
we also try to use either the overestimated or underestimated
densities as initial input values to examine the convergence.

Figure \ref{f:Fig1} shows the energy gain per site obtained by the
AF-RVB stripe states as a function of doping density in the
$t-t'-t''-J$ model with renormalized $t^{\ast}$. The reference
ground state is the uniform $d$-RVB wave function. Comparing to our
previous work \cite{CPChouPRB08}, the hopping modulation due to
electron-phonon coupling indeed stabilizes the AF-RVB stripe states
in the underdoped regime ($\delta\sim0.05-0.15$). The AF-RVB stripes
with different periods of magnetic modulation like $8a_{0}$,
$10a_{0}$, and $12a_{0}$ have its minimum at doping density $1/8$,
$1/10$ and $1/12$ respectively, which is exactly half-doped with
$1/2$ hole per charge domain. We have not found a stable stripe with
period $6a_{0}$. Above doping density greater than $1/8$, none of
the stripe states are particularly stable. These results are quite
consistent with the "Yamada plot" observed in several experiments
\cite{TranquadaNat95,KYamadaPRB98,AbbamonteNatPhys05}, except that
we have not studied the diagonal stripes below hole density of
$0.05$.  Furthermore, to make sure $8a_{0}$ is indeed most stable at
$1/8$ doping, we also calculate the energy of a "one-doped stripe"
with spin modulation $16a_{0}$ in a $16\times16$ lattice. The AF-RVB
stripe with $16a_{0}$ is found to be energetically unstable with
respect to $8a_{0}$.

\begin{figure}[top]
\rotatebox{0}{\includegraphics[height=1.5in,width=3in]{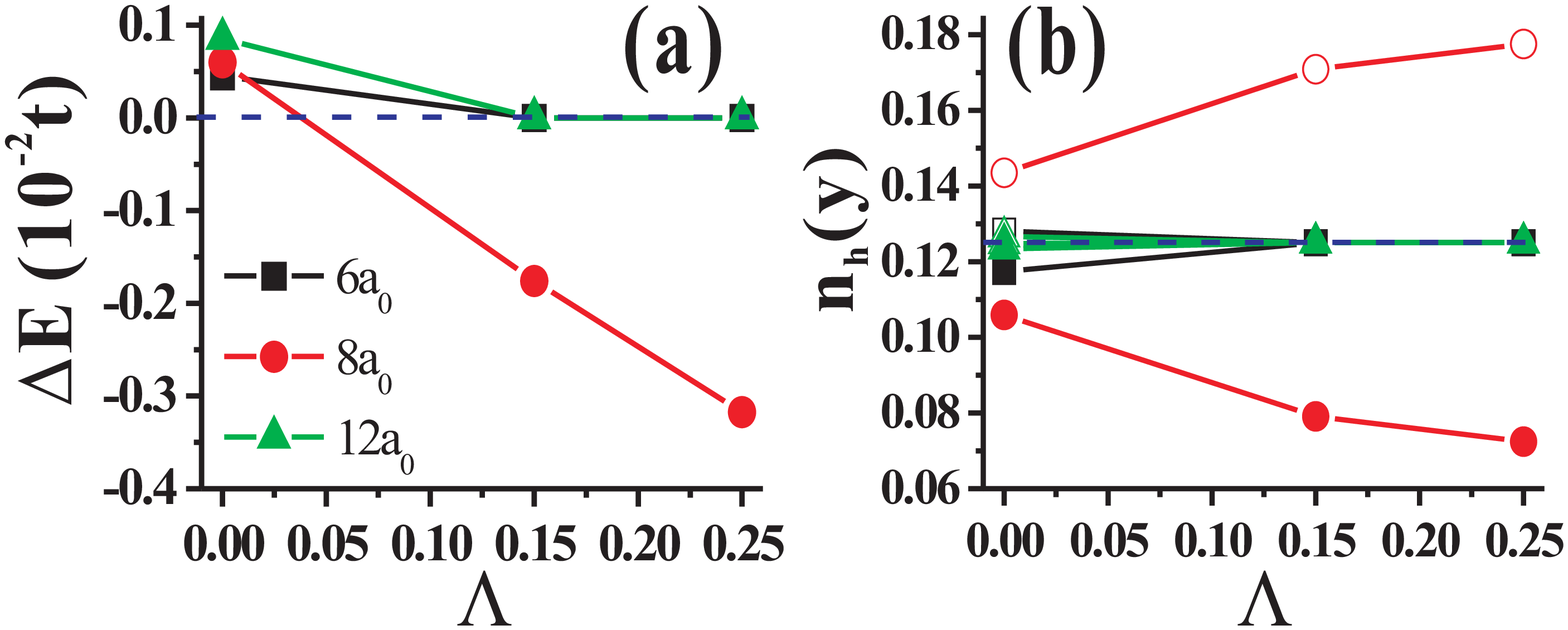}}
\caption{(Color online) (a) The optimized energy difference between
the AF-RVB stripe and uniform $d$-RVB states vs $\Lambda$. (b) The
average hole density of the AF-RVB stripes with different charge
periods vs $\Lambda$. For $6a_{0}$ and $8a_{0}$, empty [filled]
symbols indicate $n_{h}(y=1)$ [$n_{h}(y=2)$]. As for $12a_{0}$,
empty, half-filled, and filled symbols correspond to $n_{h}(y=1)$,
$n_{h}(y=2)$, and $n_{h}(y=3)$, respectively. Here the doping is
$1/8$.} \label{f:Fig2}
\end{figure}

At $1/8$ doping, Figure \ref{f:Fig2}(a) shows that in contrast with
$6a_{0}$ and $12a_{0}$, the energy gain of AF-RVB stripe state with
$8a_{0}$ increases with the strength of $\Lambda$. As shown in
Fig.\ref{f:Fig2}(b), the difference in the average hole density
between hole-rich and hole-poor domains becomes larger as $\Lambda$
increases. Among all possible magnetic periods of stripes, only the
stripe with $8a_{0}$ is stable at $1/8$ doping. We also find that
long-range pair-pair correlation function for the AF-RVB stripe
states is reduced at $1/8$ doping (not shown), which could explain
the superconducting transition temperature is suppressed near $1/8$
doping for LBCO \cite{MoodenbaughPRB88}.

\begin{figure}[top]
\rotatebox{0}{\includegraphics[height=1.5in,width=3in]{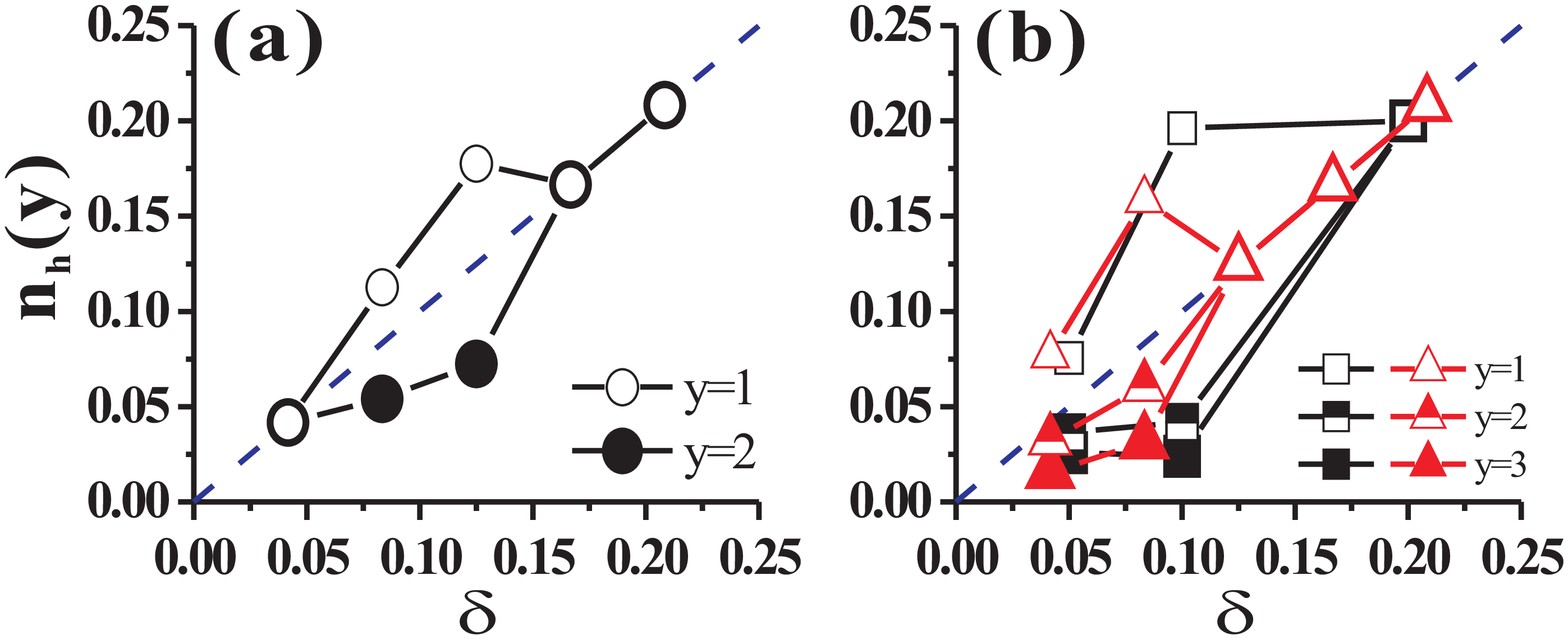}}
\caption{(Color online) The average hole density of the AF-RVB
stripes with the (a) $8a_{0}$, (b) $10a_{0}$ (squares) and $12a_{0}$
(triangles) vs the doping density $\delta$ using the same parameters
in Fig.\ref{f:Fig1}. Different symbols denote different average hole
densities along $\hat{y}$ direction. Blue dashed line is the doping
density.} \label{f:Fig3}
\end{figure}

In order to understand the stability of half-doped stripe, we
examine the spatial distribution of hole density
$n_{h}(i)(=1-\tilde{c}_{i\sigma}^{\dag}\tilde{c}_{i\sigma})$ as a
function of doping density $\delta$. In Fig.\ref{f:Fig3}(a), the
$8a_{0}$ stripes only has two distinct sites along $\hat{y}$
direction. When the difference of hole density between these two
sites is getting larger, the state gains more kinetic energy from
the modulated hopping terms as there are more holes sitting at sites
with larger $t^{\ast}$. Interestingly, the maximum energy gain (see
the filled circles in Fig.\ref{f:Fig1})) with the largest difference
in hole density exactly happens at $1/8$ doping for $8a_{0}$ stripe.
At doping density $1/6$ or higher, the kinetic energy for the
uniform state is already quite substantial, hence the separation of
hole-rich sites and hole-poor sites does not gain enough energy and
we do not find a stable periodic stripe with $8a_{0}$. This argument
also holds for stripe with other periods so there is no stable
periodic stripe state in optimum or overdoped systems. In the very
underdoped regime, since the hole density is already small,
separation into a hole-rich and a hole-poor site does not bring a
large difference to gain enough kinetic energy to stabilize the
periodic stripes. However, for stripes with $10a_{0}$ and $12a_{0}$
shown in Fig.\ref{f:Fig3}(b), there are three different sites with
only one of them having more hole density than the uniform state.
There is again the maximum difference in hole density occurs at
smaller dopings ($1/10$ and $1/12$, respectively). Similar argument
also holds for $10a_{0}$ and $12a_{0}$ stripes.

The argument provided above for the stability of half-doped stripes
could be applicable to other kinds of stripes. Hence we also examine
the site-centered AF-RVB stripes and antiphase stripes at $1/8$
doping. As expected, the site-centered stripes have similar results
as bond-centered stripes discussed above except with higher per-site
energy (0.0022t) at $8a_{0}$. Similarly for the antiphase stripes
$8a_{0}$ has lowest energy at $1/8$ doping but it again has a
slightly higher per-site energy (0.0008t) \cite{cpchou10}.

\begin{figure}[top]
\rotatebox{0}{\includegraphics[height=1.5in,width=3in]{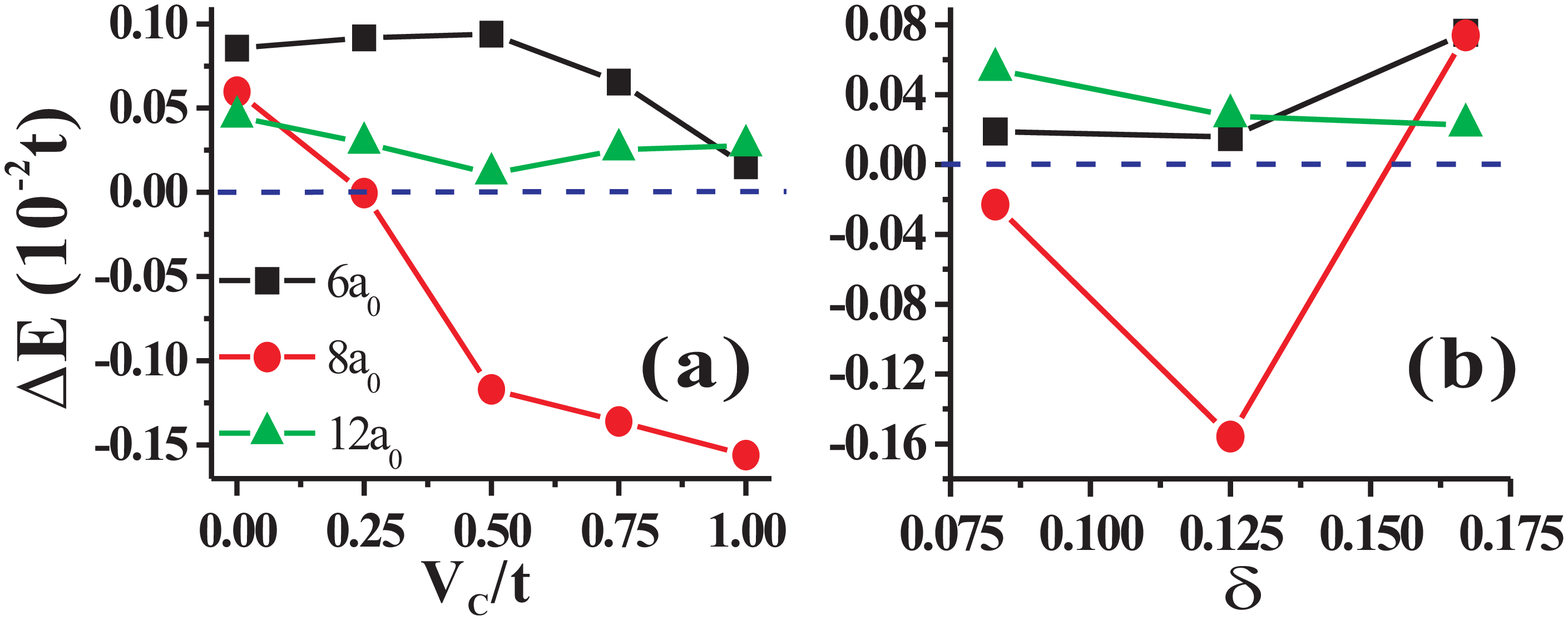}}
\caption{(Color online) (a) The optimized energy difference between
the AF-RVB stripe and uniform $d$-RVB states as a function of the
strength of Coulomb interaction $V_{C}$ at $1/8$ doping. (b) The
optimized energy difference vs doping density $\delta$ for
$V_{C}/t=1$} \label{f:Fig4}
\end{figure}

Finally, we investigate the stability of the AF-RVB stripes in
$t-t'-t''-J$ system with long-range Coulomb interaction. In this
case, the calculations omit the mass renormalization in the
Hamiltonian for the moment. The Coulomb interaction between holes is
given by
$V_{C}\sum_{i<j}n_{h}(i)n_{h}(j)/|\mathbf{r}_{i}-\mathbf{r}_{j}|$.
We have only studied stripes with periods $6a_{0}$, $8a_{0}$ and
$12a_{0}$. The modulation of hole density and staggered
magnetization is enhanced by Coulomb interaction. In
Fig.\ref{f:Fig4}(a), as $V_{C}$ increases, only the $8a_{0}$ stripe
pattern becomes more stable which is consistent with previous DMRG
studies \cite{ArrigoniPRB02}. As shown in Fig.\ref{f:Fig4}(b), the
$6a_{0}$ and $12a_{0}$ stripes always have higher energy than
uniform $d$-RVB state even for $V_{C}=t$. Although the stripe with
$8a_{0}$ is stabilized at $1/8$ doping, its energy gain is about
half of that obtained by the mass renormalization effect shown in
Fig.\ref{f:Fig2}(a). These results indicate that although the
long-range Coulomb interaction could help the formation of stripes,
it is not particularly effective in getting the half-doped stripes
as the electron-phonon coupling.

To summarize, we have successfully included the effect of mass
renormalization due to weak electron-phonon interaction with the
extended $t-J$ Hamiltonian in a self-consistent VMC calculation. It
is shown that the vertical half-doped stripes are most preferred in
the underdoped regime with hole density $1/12\leq\delta\leq1/8$.
This explains the "Yamada plot" mostly seen in LSCO materials. Since
the stability of stripes is enhanced by electron-phonon coupling,
our result also implies that this coupling is stronger in LSCO
materials than other cuprates which is consistent with the analysis
of kink in Ref.\cite{LanzaraNat01}. Additionally, we have found that
long-range Coulomb interaction does not play as an important role in
the formation of half-doped stripes as the mass renormalization
effect due to electron-phonon interaction.

This work is supported by the National Science Council in Taiwan
with Grant no.98-2112-M-001-017-MY3. The calculations are performed
in the National Center for High-performance Computing in Taiwan.


\begin{thebibliography}{99}
\bibitem{note}We roughly estimate the effective mass by considering a Holstein electron-phonon interaction as perturbation in a free Hamiltonian. By using the standard second-order perturbation theory, the effective mass can be enhanced by a factor $(1-\frac{\lambda}{\pi})^{-1}$, which is related to the electron-phonon coupling strength $\lambda$.
\bibitem{cpchou10}More details will be published in elsewhere.
\end{thebibliography}
\end{document}